\documentclass[dvipdfmx,aps,prd,preprintnumbers,superscriptaddress,nofootinbib,showpacs,twocolumn]{revtex4}%
\usepackage[dvipdfmx]{graphicx}
\usepackage{bm,latexsym,amsmath,amssymb,amsfonts,mathrsfs}
\usepackage{color}
\input{colordvi.tex}
\usepackage[dvipdfmx]{hyperref}
\usepackage{url}
\hypersetup{
    colorlinks=true,
    citecolor=cyan,
}
\newcommand*{\D}{{\rm d}}
\newcommand*{\mpl}{M_{\rm Pl}}
\begin{document}

\title{%Completing Inflation with Galilean Genesis
%Galilean Genesis of the Inflationary Universe
Galilean Creation of the Inflationary Universe}

\author{Tsutomu~Kobayashi}
\email[Email: ]{tsutomu"at"rikkyo.ac.jp}
\affiliation{Department of Physics, Rikkyo University, Toshima, Tokyo 175-8501, Japan}

\author{Masahide~Yamaguchi}
\email[Email: ]{gucci"at"phys.titech.ac.jp}
\affiliation{Department of Physics, Tokyo Institute of Technology, Tokyo 152-8551, Japan}

\author{Jun'ichi~Yokoyama}
\email[Email: ]{yokoyama"at"resceu.s.u-tokyo.ac.jp}
\affiliation{Research Center for the Early Universe (RESCEU),Graduate School of Science, The University of Tokyo, Tokyo 113-0033, Japan }
\affiliation{Department of Physics, Graduate School of Science, The University of Tokyo, Tokyo 113-0033, Japan}
\affiliation{Kavli Institute for the Physics and Mathematics of the Universe (Kavli IPMU), UTIAS, WPI, The University of Tokyo, Kashiwa, Chiba 277-8568, Japan}

\begin{abstract}
It has been pointed out that the null energy condition can be violated stably
in some non-canonical scalar-field theories.
This allows us to consider the Galilean Genesis scenario in which
the universe starts expanding
from Minkowski spacetime and hence is free from the initial singularity.
We use this scenario to study the early-time completion of inflation,
pushing forward the recent idea of Pirtskhalava {\em et al}.
We present a generic form of the Lagrangian governing the background and perturbation
dynamics in the Genesis phase, the subsequent inflationary phase,
and the graceful exit from inflation, as opposed to employing the effective
field theory approach.
Our Lagrangian belongs to a more general class of
scalar-tensor theories than the Horndeski theory and
Gleyzes-Langlois-Piazza-Vernizzi generalization, but still has the same number
of the propagating degrees of freedom, and thus can avoid Ostrogradski instabilities.
We investigate the generation and evolution of primordial perturbations
in this scenario and show that one can indeed construct a stable model of inflation
preceded by (generalized) Galilean Genesis.
\end{abstract}

\pacs{98.80.Cq}
\preprint{RUP-15-6, RESCEU-10/15}
\maketitle
%%%%%%%%%%%%%%%%%%%%%%%%%%%%%%%%%%%%%%%%%%%%%%%%%%%%%%%%%%%%%%%%%%%%%%

\section{Introduction}

Inflation in the early Universe \cite{inflation,Yokoyama:2014nga}
is now an indispensable ingredient of
modern cosmology not only to explain the global properties of
homogeneous and isotropic space with a vanishingly small spatial
curvature but also to account for the origin of the primordial curvature
perturbation that seeded cosmic structure formation \cite{fluctuation}.
At present, despite the significant progress in the state-of-the-art
precise measurements of the cosmic microwave background radiation (CMB)
by WMAP~\cite{Bennett:2012zja,Hinshaw:2012aka}
and Planck~\cite{Ade:2013ktc,Ade:2013uln} missions, there is no single
observational result in conflict with the single-field inflation
paradigm \cite{Yokoyama:2014nga}.
In particular, the
anti-correlation of the temperature and the E-mode polarization
anisotropies on large scales observed by the WMAP mission strongly
supports the superhorizon perturbations suggested by 
inflation \cite{Peiris:2003ff}.

In other words, once inflation sets in, virtually all the available
cosmological observation data  can be explained simultaneously
irrespective of the initial condition of the Universe.
This does not mean that we may be indifferent to the initial condition
of the Universe before inflation.  On the contrary, in order to achieve
complete understanding of the cosmic history, we must work out the very
beginning of the Universe that may smoothly evolve into the inflationary
phase.  

 As is well known, as long
as the null energy condition (NEC) is satisfied in the expanding phase, the Hubble parameter and
the energy density of the universe increase backward in cosmic time.
So, it is often claimed that, if one tries to discuss what
happened before inflation and/or how inflation started, one needs to
know the information of very high energy physics, and challenge the
initial singularity problem \cite{Borde:1996pt} in terms of quantum gravity. 
But, this is not always the case.

Recently, it was recognized that, if an action includes higher derivative
terms of a scalar field like the Galileon terms, the NEC can be violated without
encountering ghost nor gradient instabilities. See,
e.g., Ref.~\cite{Rubakov:2014jja} for a recent review and
Ref.~\cite{Sawicki:2012pz} for a subtle issue of nonlinear instabilities.
If the NEC is violated, the energy density can grow as time proceeds,
contrary to the conventional wisdom. 
In the NEC violating theories, the universe can therefore
start from the static zero-energy state described by
the Minkowski spacetime from infinite past~\cite{Creminelli:2006xe},
and the universe starts expansion with the increase of the energy
density.

Such a picture of the emergence of the universe was first proposed by
Creminelli {\it et al.}~\cite{Creminelli:2010ba} with the name 
Galilean Genesis.
In their model, however, the hot big bang state was postulated to be
realized after the effective field theory description breaks down as the
energy density blows up beyond its realm of validity.  Therefore, the theory
to describe the most important epoch of the early universe is lacking
there.  

Nevertheless, since their original idea is so interesting that a number
of extension has been made in 
a wider class of scalar field theories~\cite{Creminelli:2012my,Hinterbichler:2012fr,Hinterbichler:2012yn,Nishi:2014bsa,Nishi:2015pta}
and various aspects of the Genesis scenario have been explored in
the literature
\cite{LevasseurPerreault:2011mw,Wang:2012bq,Easson:2013bda,Rubakov:2013kaa,Elder:2013gya},
such as avoidance of the superluminal propagation of perturbations
and absence of primordial tensor perturbations.
They have been unsuccessful, however, to realize transition from the
Genesis phase to the hot big bang state
within their model Lagrangians.  

In this paper,  we take a different
approach, namely, to make use of the Galilean Genesis to explain the
initial condition of the Universe before inflation and smoothly connect
it to the inflationary phase, thereby solving the initial singularity
problem 
\cite{Borde:1996pt} and the 
trans-Planckian problem \cite{Martin:2000xs} 
(see also \cite{Starobinsky:2001kn}) in inflationary cosmology.

In fact, such an approach has also been put forward by Pirtskhalava
{\it et al.}~\cite{Pirtskhalava:2014esa} recently. 
Their model Lagrangian, however, gives rise to gradient 
instability as it is, although it has been argued there that higher-order structure 
of the effective field theory for perturbations possesses enough freedom to 
cure the gradient instability. 
Discussion on termination of inflation and reheating is not presernted there, either.

In the present paper, we construct
a specific model free from any catastrophic instabilities
and with subluminal velocities of primordial perturbations. In our setup
the universe starts from the Minkowski spacetime from infinite
past and is smoothly connected to the 
inflationary phase followed by the graceful
exit. 
For this purpose, we provide a generic Lagrangian capable of
describing the background and perturbation evolution in all the above
phases instead of choosing the effective field theory approach because
the latter cannot capture the evolution of the background and
perturbations from pre-inflationary Genesis to the exit from inflation
with the same single Lagrangian.

Although we start with asymptotically Minkowski space at the past
infinity for aesthetic beauty, it has been shown that the Galilean
Genesis solution is an attractor for a variety of initial conditions
including those with a negative Hubble parameter and/or finite
curvature, provided that the time derivative of the scalar field has the
right sign \cite{Nishi:2015pta}.

The Horndeski theory~\cite{Horndeski}
or the generalized Galileon \cite{Deffayet:2011gz}, whose mutual
equivalence was first shown in \cite{Kobayashi:2011nu},
is known to be the
most general scalar-tensor tensor theory with the second-order
field equations, and thereby avoid Ostrogradski instabilities in spite of having
higher derivative terms in the action.
%But, in order to avoid the
%instabilities, only time derivatives in equations of motion should be
%second order while spacial derivatives can include high
%derivatives.
The theory can be generalized to have second-order field equations
only in a specific gauge while maintaining the number of propagating degrees of freedom.
This possibility was realized recently by
Gleyzes {\it et al.}~\cite{GLPV1} (see also Ref.~\cite{Zumalacarregui:2013pma}) and was extended
further by Gao~\cite{Gao:2014soa}.
The number of
propagating degrees of freedom in these theories
is indeed shown to be the same as that of the
Horndeski theory~\cite{GLPV1,Gao:2014soa,Lin:2014jga,GLPV2,Gao:2014fra,Fasiello:2014aqa}.
In this paper,
we use the subclass of Gao's framework as a concrete realization of the unified
scenario starting from Galilean Genesis through inflation to the graceful exit.

This paper is organized as follows. In the next section, we give a
framework of our model and derive the background equations of motion and
the quadratic actions of cosmological perturbations. In Sec.~III,
a concrete Lagrangian is constructed to describe our scenario beginning from the
Genesis phase through the inflationary one to the graceful exit, and such a background
dynamics is presented explicitly. In Sec.~IV, we discuss the stability
during each phase based on the quadratic actions of cosmological
perturbations. In Sec.~V, a concrete realization of our scenario
is given. The final section is devoted to our conclusions and
discussion.

\section{General Framework}

Let us start with describing the general framework to construct and
study our explicit realization of the early-time completion of
inflation.  We would like to consider theories composed of a metric
$g_{\mu\nu}$ and a single scalar field $\phi$, and hence it will be
appropriate to work in the Horndeski theory.  The Lagrangian of the
Horndeski theory is of the form
\begin{eqnarray}
{\cal L}&=&\sqrt{-g}
\bigl[G_2(\phi, X)-G_3(\phi, X)\Box\phi + G_4(\phi, X)R^{(4)}
\nonumber\\&&
+G_5(\phi, X)G_{\mu\nu}^{(4)}\nabla^\mu\nabla^\nu\phi
+\cdots\bigr],\label{Horndeski-L}
\end{eqnarray}
where $X:=-g^{\mu\nu}\partial_\mu\phi\partial_\nu\phi/2$,
$R^{(4)}$ is the four-dimensional Ricci scalar,
and $G_{\mu\nu}^{(4)}$ is the four-dimensional Einstein tensor.
We have four arbitrary functions of $\phi$ and $X$ in the Horndeski theory.
This is the most general Lagrangian having second-order field equations.
Nevertheless, it will turn out that this framework is insufficient for our purpose,
and hence we have to go beyond the Horndeski theory.

One can generalize the Horndeski theory to possess higher order field
equations while maintaining the number of propagating degrees of
freedom~\cite{GLPV1}.  The first step to do so is to perform an ADM
decomposition by taking $\phi=$ const hypersurfaces as constant time
hypersurfaces.  In the ADM language, the metric is written as
\begin{eqnarray}
\D s^2=-N^2\D t^2+\gamma_{ij}\left(\D x^i+N^i\D t\right)\left(\D x^j+N^j\D t\right).
\end{eqnarray}
By definition $\phi$ is a function of only $t$, $\phi=\phi(t)$, and 
$X=\dot\phi^2/2N^2$,
where a dot denotes differentiation with respect to $t$, so any function
of $\phi$ and $X$ can be regarded as a function of $t$ and the lapse
function $N$, provided that $\dot\phi$ and $N^{-1}$ never vanish.
Then, the Horndeski Lagrangian~(\ref{Horndeski-L})
can be written in terms of the ADM variables as
${\cal L}=\sqrt{\gamma}N\sum_a L_a$ with
\begin{eqnarray}
&&L_2=A_2(t,N),
\quad
L_3=A_3(t,N)K,
\nonumber\\
&&
L_4=A_4(t,N)\left( K^2- K_{ij}^2\right)+B_4(t,N)R,
\nonumber\\
&&
L_5=A_5(t,N)\left( K^3-3 KK_{ij}^2
+2K_{ij}^3\right)
\nonumber\\&&
\qquad\quad
+B_5(t,N)K^{ij}\left(R_{ij}-\frac{1}{2}g_{ij}R\right),
\end{eqnarray}
where $K_{ij}$ and $R_{ij}$ are the extrinsic and intrinsic curvature tensors
on the constant time hypersurfaces, and $A_4$, $A_5$, $B_4$, and $B_5$
are subject to the relations
\begin{eqnarray}
A_4=-B_4-N\frac{\partial B_4}{\partial N},
\quad
A_5=\frac{N}{6}\frac{\partial B_5}{\partial N}.
\label{AB-constraint}
\end{eqnarray}
Variation of the above Lagrangian with respect to $N$
gives a second-class constraint that eliminates only one degree of freedom,
as opposed to general relativity.
The key trick to generalize the Horndeski theory is to notice
that this property remains the same even if one liberates $A_4$ and $A_5$
from the restriction imposed by Eq.~(\ref{AB-constraint})~\cite{GLPV1}.
We thus arrive at the so called GLPV theory that is more general than Horndeski
but has the same number of propagating degrees of freedom.
One can move back to a covariant form of the Lagrangian by introducing the unit normal
to the constant time hypersurfaces as $n_\mu = -\partial_\mu\phi/\sqrt{2X}$,
writing the extrinsic curvature tensor in terms of $n_\mu$,
and using the Gauss-Codazzi equations.
Since there are six arbitrary functions of $t$ and $N$ in the ADM form,
the resultant covariant Lagrangian has six arbitrary functions of $\phi$ and $X$.

The above idea has been pushed forward by 
Gao~\cite{Gao:2014soa}, who proposed a unified framework to study
single scalar-tensor theories beyond Horndeski.  One can write a general
Lagrangian in the ADM form as
\begin{eqnarray}
{\cal L}&=&\sqrt{\gamma}N\bigl[d_0+d_1R+d_2R^2+\cdots
+\left(a_0+a_1R+\cdots\right)K
\nonumber\\&&
+\left(a_2R^{ij}+\cdots\right)K_{ij}+b_1K^2+b_2K_{ij}K^{ij}+\cdots
\bigr],
\end{eqnarray}
where the coefficients $d_0$, $d_1$, ... are arbitrary functions of $t$
and $N$.  The Hamiltonian depends nonlinearly on $N$ as in the GLPV
theory, giving rise to a single scalar degree of freedom on top of the
traceless and transverse gravitons~\cite{Gao:2014fra}.

In this paper, we will employ the Lagrangian ${\cal
L}=\sqrt{\gamma}N\sum_a L_a$ with
\begin{eqnarray}
&&L_2=A_2(t,N),
\quad
L_3=A_3(t,N)K,
\nonumber\\
&&
L_4=A_4(t,N)\left(\lambda_1 K^2- K_{ij}^2\right)+B_4(t,N)R,
\nonumber\\
&&
L_5=A_5(t,N)\left(\lambda_2 K^3-3\lambda_3 KK_{ij}^2
+2K_{ij}^3\right)
\nonumber\\&&
\qquad \quad
+B_5(t,N)K^{ij}\left(R_{ij}-\frac{1}{2}g_{ij}R\right),\label{ourLag}
\end{eqnarray}
where $\lambda_1$, $\lambda_2$, and $\lambda_3$ are constant parameters
of the theory. This is a deformation of the GLPV Lagrangian and belongs
to a subclass of Gao's framework.  The generalization to this level is
sufficient for the purpose of the present paper.  The GLPV theory is
recovered by taking $\lambda_1=\lambda_2=\lambda_3=1$.

Given the Lagrangian~(\ref{ourLag}) in the ADM form, one can restore the
scalar degree of freedom $\phi$ to write its covariant expression in the
same way as in the GLPV theory.  However, it will be more convenient for
our purpose to use the explicitly time-dependent Lagrangian, because by
doing so one can easily design the Lagrangian so as to admit the desired
cosmological evolution.

Before specifying the suitable form of $A_2(t,N)$, $A_3(t,N)$, ... to
construct our early universe model, let us derive the general equations
governing the background and perturbation dynamics of cosmologies based
on the Lagrangian~(\ref{ourLag}).  The ADM variables are given by
\begin{eqnarray}
&&
N= \overline{N}(t)\left(1 + \delta n\right),
\quad
N_i=\overline{N}\partial_i \chi,
\nonumber\\&&
\gamma_{ij} = a^2(t)e^{2\zeta}\left(\delta_{ij}+h_{ij}+\frac{1}{2}h_{ik}h_{kj}\right),
\label{ADM-cosmological}
\end{eqnarray}
where $\zeta$ is the curvature perturbation in the unitary gauge and
$h_{ij}$ is the transverse and traceless tensor perturbation.  A
spatially flat background has been assumed and the spatial
diffeomorphism invariance was used to write $\gamma_{ij}$ in the above
form.  In the following, the background value of the lapse function is
denoted by $N$ where there is no worry about confusion.

\subsection{Background Equations}

Substituting Eq.~(\ref{ADM-cosmological}) to the
Lagrangian~(\ref{ourLag}), we obtain the background part of the
Lagrangian as
\begin{eqnarray}
{\cal L}^{(0)}&=&Na^3\left(A_2+3A_3H+6\eta_4A_4H^2+6\eta_5A_5H^3\right),\label{zero-bg}
\end{eqnarray}
where $\eta_4:=(3\lambda_1-1)/2$, $\eta_5:=(9\lambda_2-9\lambda_3+2)/2$,
and $H:=\dot a/(Na)$.
At the background level, $\lambda_1$, $\lambda_2$, and $\lambda_3$
just rescale $A_4$ and $A_5$. In what follows we simply consider the case with $\eta_4>0\;\Leftrightarrow\; \lambda_1>1/3$.
Since we are considering a spatially flat
universe, we have $R_{ij}=0$ at zeroth order, and hence $B_4$ and $B_5$ play no
role in the background dynamics.  Varying Eq.~(\ref{zero-bg}) with
respect to $N$ and $a$, we obtain, respectively,
\begin{eqnarray}
-{\cal E}&:=&(NA_2)'+3NA_3'H+6\eta_4N^2(N^{-1}A_4)'H^2
\nonumber\\&&
+6\eta_5 N^3(N^{-2}A_5)'H^3
\nonumber\\
&=&0,
\\
{\cal P}&:=&A_2-6\eta_4A_4H^2-12\eta_5A_5H^3
\nonumber\\&&
-\frac{1}{N}\frac{\D}{\D t}\left(
A_3+4\eta_4A_4H+6\eta_5A_5H^2
\right)
\nonumber\\&=&0,
\end{eqnarray}
where a prime represents differentiation with respect to $N$.
The background equations contain at most second derivatives of the scale
factor and first derivatives of the Lapse function.

\subsection{Cosmological Perturbations}\label{sec:CP}

The quadratic Lagrangian for the tensor perturbation is given by
\begin{eqnarray}
{\cal L}^{(2)}_{T}=\frac{Na^3}{8}\left[
\frac{{\cal G}_T}{N^2}\dot h_{ij}^2
-\frac{{\cal F}_T}{a^2}(\partial h_{ij})^2\right],
\end{eqnarray}
where
\begin{eqnarray}
{\cal G}_T&:=&-2A_4-6\left(3\lambda_3-2\right)A_5H,
\\
{\cal F}_T&:=&2B_4 +\frac{1}{N}\frac{\D B_5}{\D t}.
\end{eqnarray}
The equation of motion contains at most second derivatives both in time and space.
The tensor perturbation is stable provided that ${\cal G}_T>0$
and ${\cal F}_T>0$.

The quadratic Lagrangian for the scalar perturbations is given by
\begin{eqnarray}
{\cal L}^{(2)}_S&=&Na^3
\Biggl[
-3{\cal G}_A\frac{\dot\zeta^2}{N^2}+\frac{{\cal F}_T}{a^2}(\partial\zeta)^2
+\Sigma \delta n^2
\nonumber\\&&\qquad
-2\Theta\delta n\frac{\partial^2\chi}{a^2}
+2{\cal G}_A\frac{\dot\zeta}{N}\frac{\partial^2\chi}{a^2}
+6\Theta\delta n\frac{\dot\zeta}{N}
\nonumber\\&&\qquad
-2{\cal G}_B\delta n\frac{\partial^2\zeta}{a^2}
-{\cal C}\frac{(\partial^2\chi)^2}{a^4}
\Biggr],\label{LagS2}
\end{eqnarray}
where the coefficients are defined as
\begin{eqnarray}
\Sigma&:=&NA_2'+\frac{1}{2}N^2A_2''+\frac{3}{2}N^2A_3''H
\nonumber\\&&
+3\eta_4\left(2A_4-2NA_4'+N^2A_4''\right)H^2
\nonumber\\&&
+3\eta_5\left(6A_5-4NA_5'+N^2A_5''\right)H^3,
\\
\Theta&:=&\frac{NA_3'}{2}-2\eta_4\left(A_4-NA_4'\right)H
\nonumber\\&&
-3\eta_5\left(2A_5-NA_5'\right)H^2,
\\
{\cal G}_A&:=&-2\eta_4A_4-6\eta_5A_5H,
\\
{\cal G}_B&:=&2\left(B_4+NB_4'\right)-HNB_5',
\\
{\cal C}&:=&(1-\lambda_1)A_4-(6+9\lambda_2-15\lambda_3)A_5H,
\end{eqnarray}
and note the relation ${\cal G}_T={\cal G}_A-3{\cal C}$.
One has ${\cal C}=0$ in the Horndeski and GLPV theories,
in which $\lambda_1=\lambda_2=\lambda_3=1$.
Therefore, the last term in the Lagrangian~(\ref{LagS2})
is the novel consequence of theories beyond GLPV.

From $\delta{\cal L}_S^{(2)}/\delta(\delta n)=0$ and
$\delta{\cal L}_S^{(2)}/\delta(\partial^2\chi)=0$ we obtain
\begin{eqnarray}
\delta n&=&\frac{1}{\Theta^2+\Sigma{\cal C}}\left[\Theta
({\cal G}_A-3{\cal C})\frac{\dot\zeta}{N}+{\cal G}_B{\cal C}\frac{\partial^2\zeta}{a^2}
\right],\label{ELdeltan}
\\
\frac{\partial^2\chi}{a^2}
&=&\frac{1}{\Theta^2+\Sigma{\cal C}}\left[
(3\Theta^2+\Sigma{\cal G}_B)\frac{\dot\zeta}{N}
-\Theta{\cal G}_B\frac{\partial^2\zeta}{a^2}
\right].\label{ELchi}
\end{eqnarray}
Substituting Eqs.~(\ref{ELdeltan}) and~(\ref{ELchi}) into
Eq.~(\ref{LagS2}), we obtain
the reduced Lagrangian for the curvature perturbation,
\begin{eqnarray}
{\cal L}^{(2)}_S =Na^3\left[
{\cal G}_S\frac{\dot\zeta^2}{N^2}+\zeta\left({\cal F}_S\frac{\partial^2}{a^2}
-{\cal H}_S\frac{\partial^4}{a^4}\right)\zeta\right],
\end{eqnarray}
where
\begin{eqnarray}
{\cal G}_S&:=&\frac{\Sigma{\cal G}_T^2}{\Theta^2+\Sigma{\cal C}}+3{\cal G}_T,
\\
{\cal F}_S&:=&\frac{1}{Na}\frac{\D}{\D t}\left(
\frac{a\Theta{\cal G}_B{\cal G}_T}{\Theta^2+\Sigma{\cal C}}
\right)-{\cal F}_T,
\\
{\cal H}_S&:=&\frac{{\cal G}_B^2{\cal C}}{\Theta^2+\Sigma{\cal C}}.
\end{eqnarray}
Thus, if ${\cal C}\neq 0$, the equation of motion for $\zeta$ has the fourth derivative in space,
giving the dispersion relation
\begin{eqnarray}
\omega^2=\frac{{\cal F}_S}{{\cal G}_S}k^2+\frac{{\cal H}_S}{{\cal G}_S}\frac{k^4}{a^{2}}.
\end{eqnarray}
We require that ${\cal G}_S>0$ in order to avoid ghost instabilities.
However, we allow for a negative sound speed squared, $c_s^2:={\cal F}_S/{\cal G}_S<0$,
for a short period of time.
In the absence of the $k^4$ term (${\cal C}=0$), a negative sound speed squared would
cause a rapid growth of instabilities for large $k$ modes.
In this paper, we consider theories with ${\cal C}\neq 0$, so that
the curvature perturbation with large $k$
can be stabilized by requiring that ${\cal H}_S/{\cal G}_S>0$.

As will be seen in the rest of the paper,
the sound speed squared becomes negative at the transition
from one phase to another. 
Such a behavior should not occur even for a tiny period because
high wavenumber modes would grow exponentially rapidly.
However, we could not avoid it not only within the Horndesky theory
but also the GLPV theory despite we analyzed extensive models.
On the other hand, we have not been successful in proving that
this is an inevitable consequence.
Since our primary purpose is to show an existence proof of the
model to realize our intended cosmic evolution without any
instabilities, we construct
a specific model by  going beyond the GLPV theory and invoking the $k^4$ term.

\section{Starting Inflation from Minkowski}

\subsection{Construction of the Lagrangian}

The Lagrangian we study in this paper is characterized by
a single time-dependent function $f(t)$
and four functions $a_2, a_3, a_4, a_5$ of $N$:
\begin{eqnarray}
A_2&=&M^4_{2}f^{-2(\alpha +1)}a_2(N),  \label{27}
\\
A_3&=&M^3_{3}f^{-(2\alpha+1)}a_3(N),
\\
A_4&=&-\frac{\mpl^{2}}{2}+M^2_{4}f^{-2\alpha} a_4(N),
\\
A_5&=&M_{5}f a_5(N), \label{30}
\end{eqnarray}
where $\alpha \;(>0)$ is a constant parameter.
We have introduced the mass scales $M_a$ (and the Planck mass $\mpl$),
so that $f(t)$ and $a_a(N)$ are dimensionless.
The other two functions, $B_4$ and $B_5$, are arbitrary at this stage
because they have no impacts on the background dynamics.
Note that $f$ is not a dynamical variable.
Specifying the functions $f=f(t)$ and $a_a=a_a(N)$
amounts to defining a concrete theory.
The above forms of $A_a$ are
chosen so that the theory admits an inflationary universe 
preceded by the generalized Galilean Genesis
while retaining much of the generality.
Other choices could be possible
and hence we do not claim that this is the most general
 description of such scenarios at all.
Instead, as we mentioned above, we would provide the existence
proof of desired models by demonstrating that a sufficiently wide class
 of healthy models can indeed be constructed.

We design $f(t)$ so as to implement
the (generalized) Galilean Genesis followed by inflation
and a graceful exit from the prolonged inflationary phase.
Our choice is
\begin{eqnarray}
f \approx \dot f_0 t
\quad(\dot f_0= {\rm const} <0)\label{f:gen}
\end{eqnarray}
well before $t=t_0$, and
\begin{eqnarray}
f \simeq f_1={\rm const}
\label{f:dS}
\end{eqnarray}
for $t\gtrsim t_0$.
As our time variable starts at $t=-\infty$ with asymptotically Minkowski
spacetime configuration, $t$ is large and negative in the beginning, so 
 we find  $f\gg 1$ in Eq.~(\ref{f:gen}).
As will be seen shortly,
the initial stage described by Eq.~(\ref{f:gen})
corresponds to the generalized Galilean Genesis,
while the subsequent stage described by Eq.~(\ref{f:dS})
to inflation.
After a sufficiently long period of the inflationary stage, we assume that
\begin{eqnarray}
f \sim t^{1/(\alpha +1)}
\label{f:exit}
\end{eqnarray}
for $t\gtrsim t_{\rm end}$, where $t_{\rm end}$ is the time at the end of inflation.
With this the universe exits from inflation.
In what follows we will investigate the background evolution of each stage.

\subsection{Genesis Phase}

Assuming that $H\sim |t|^{-(2\alpha +1)}$ in the first stage
where $f$ is given by Eq.~(\ref{f:gen}),
let us look for a consistent solution for large $f$.
The background field equations read
\begin{eqnarray}
-{\cal E}&=&M^4_{2}f^{-2(\alpha+1)}(Na_2)'+{\cal O}(f^{-4\alpha-2})=0,
\label{Fr:gen}
\\
{\cal P}&=&
-\frac{1}{N}\frac{\D}{\D t}
\left(M^3_{3}f^{-(2\alpha+1)}a_3-2\eta_4\mpl^2H\right)
\nonumber\\&&+M^4_{2}f^{-2(\alpha+1)}a_2+{\cal O}(f^{-4\alpha-2})=0.
\label{Pres:gen}
\end{eqnarray}
It can be seen from Eq.~(\ref{Fr:gen}) that
the lapse function $N$ is a constant, $N=N_0$, satisfying
\begin{eqnarray}
a_2(N_0)+N_0a_2'(N_0)=0.
\end{eqnarray}
Then, $H$ is consistently determined from Eq.~(\ref{Pres:gen}),
which can be written as
\begin{eqnarray}
\frac{2\eta_4\mpl^2}{N_0}\frac{\D H}{\D t}+f^{-2(\alpha+1)}\hat p=0,
\end{eqnarray}
where
\begin{eqnarray}
\hat p&=&
M_{2}^{4}a_2(N_0)+(2\alpha +1)M_{3}^{3}a_3(N_0)\frac{\dot f_0}{N_0}
\end{eqnarray}
is a constant.
This leads to the generalized Galilean Genesis solution~\cite{Nishi:2015pta}:
\begin{eqnarray}
H&=&  -\frac{\hat p}{2(2\alpha +1)\eta_4 \mpl^2} \frac{N_0}{|\dot f_0|}f^{-(2\alpha +1)}
\sim \frac{1}{(-t)^{2\alpha +1}},
\\
a&=&1- \frac{\hat p}{4\alpha(2\alpha +1)\eta_4 \mpl^2} \frac{N_0^2}{\dot f_0^2}f^{- 2\alpha }.
\end{eqnarray}
It is required that $\hat p/\eta_4<0$ to guarantee $H>0$.
We have thus arrived at the generalized Galilean Genesis solution
starting from the Lagrangian written in the ADM form
rather than in the covariant form.
The original Galilean Genesis solution found in Ref.~\cite{Creminelli:2010ba}
corresponds to $\alpha = 1$.

In deriving the above solution,
$M_4^2f^{-2\alpha}a_4\;\left(\subset A_4\right)$ and $M_5fa_5\;\left(=A_5\right)$
are always subdominant due to the assumed scalings $\sim f^{-2\alpha}$ and $\sim f$.
Therefore, any choices of $a_4(N)$ and $a_5(N)$ will not spoil
the above Galilean Genesis solution.
As will be seen in the next section, those two terms are also irrelevant
to the stability conditions during the Genesis phase.

\subsection{Inflationary Phase}

The Galilean Genesis phase will end at $t\sim t_0$
since the function $f$ is constant for $t\gtrsim t_0$.
In the subsequent phase we obtain the de Sitter solution,
$N=N_{\rm inf}=$ const and $H=H_{\rm inf}=$ const,
satisfying
\begin{eqnarray}
-{\cal E}&=&(N_{\rm inf}A_2)'+3N_{\rm inf}A_3'H_{\rm inf}+6\eta_4N^2_{\rm inf}(N_{\rm inf}^{-1}A_4)'H_{\rm inf}^2
\nonumber\\
&&
+6\eta_5 N_{\rm inf}^3(N_{\rm inf}^{-2}A_5)'H^3_{\rm inf}=0,
\\
{\cal P}&=&
A_{2}-6\eta_{4}A_{4}H_{\rm inf}^{2}-12\eta_{5}A_{5}H_{\rm inf}^{3}=0.
\end{eqnarray}
(Note that $A_a$ is now a function of $N$ only and is independent of $t$.)

A $t$-independent Lagrangian in the ADM form can be recast in
a covariant Lagrangian with the shift symmetry, $\phi\to\phi + c$.
This implies that the above exact de Sitter solution corresponds to
kinetically driven G-inflation.
If one invokes a weak time-dependence in $f$,
one obtains quasi-de Sitter inflation instead.

\subsection{Graceful Exit}

After the prolonged phase of inflation,
$f$ is given by Eq.~(\ref{f:exit}).
We assume that $t$ is sufficiently large, so that $f\gg 1$.
Then, we have a consistent solution with $N=N_{\rm e}=$ const and
$H^2\sim 1/t^2\sim f^{-2(\alpha +1)}\sim A_2$
satisfying
\begin{eqnarray}
-{\cal E}&=&(N_{\rm e}A_{2})'+3\eta_{4}\mpl^{2}H^{2}+{\cal O}(f^{-(3\alpha + 2)})
\nonumber\\&&
=0,\label{exit:eq1}
\\
{\cal P}&=&A_{2}+3\eta_{4}\mpl^{2}H^{2}+\frac{2\eta_{4}\mpl^2}{N_{\rm e}}\frac{\D H}{\D t}
+{\cal O}(f^{-(3\alpha + 2)})
\nonumber\\&&
=0.\label{exit:eq2}
\end{eqnarray}
Thus, one can implement a graceful exit from inflation.
It follows from Eq.~(\ref{exit:eq1}) that
\begin{eqnarray}
(N_{\rm e}a_2)'<0.\label{exit:c1}
\end{eqnarray}
It can be shown using Eqs.~(\ref{exit:eq1}) and~(\ref{exit:eq2}) that,
during this third stage,
\begin{eqnarray}
H^{2}\propto \frac{1}{a^{m}},
\quad
m:=\frac{3N_{\rm e}a_{2}'}{(N_{\rm e}a_{2})'}.
\end{eqnarray}
It is therefore necessary to impose $m>0\;\Leftrightarrow\; a_2'<0$.

In the standard potential-driven inflation models \cite{Yokoyama:2014nga}
inflation is followed by coherent field oscillation of the inflaton
scalar field which decays to radiation to reheat the universe.
In the present approach the scalar field $\phi$ is used to specify
constant time hypersurfaces, so that $\dot\phi$ may not vanish in order
to preserve one-to-one correspondence between $\phi$ and the cosmic time
$t$.  Hence one must switch from the ADM language we used to construct
the action to the conventional ``$\phi$ language'' at this point in
order to apply the standard reheating mechanism, which is all right but
looks like sewing a fox's skin to the lion's.

Here instead we consider another reheating mechanism which can take
place without breaking the one-to-one correspondence between $\phi$
and $t$, namely, the gravitational reheating due to the change of
geometry or the cosmic expansion
law~\cite{Parker:1968mv,Parker:1969au,Zeldovich:1971mw,Birrell:1979iq,Ford:1986sy}.

During the transition from the de Sitter inflation to a decelerated
power-law expansion, conformally non-invariant particles are produced
with
the initial energy density
\begin{equation}
 \rho_r =\sigma H_{\rm inf}^4,
\end{equation}
where $\sigma$ is a factor determined by the effective number of conformally
noninvariant fields and the change of the geometry.  For example, for
$m=6$
or 4, a
single minimally coupled massless scalar field contributes to $\sigma$
by
\begin{align}
  \sigma_1&=\frac{9}{32\pi^2}\ln\left(\frac{1}{H\Delta t}\right),&~~ (m=6), \\
  \sigma_1&=\frac{1}{8\pi^2}\ln\left(\frac{1}{H\Delta t}\right),&~~ (m=4), 
\end{align}
respectively \cite{Ford:1986sy,Kunimitsu:2012xx}.  Here $\Delta t$ is
the time required for the transition.
In case it is
nonminimally coupled with a coupling parameter $\xi$, a factor
$(1-6\xi)^2$
is multiplied there.

In order for the radiation thus created to dominate the universe,
the energy density of the scalar field must dissipate more rapidly, namely,
\begin{eqnarray}
m>4\;\;\;\Leftrightarrow\;\;\;4a_2+N_{\rm e}a_2'>0,\label{exit:c2}
\end{eqnarray}
then, the reheating temperature at the radiation domination is given by
\begin{equation}
 T_R=\left(\frac{30}{\pi^2g_\ast}\right)^{1/4}
\left(\frac{\sigma^{m/4}}{3}\right)^{1/(m-4)}
\left(\frac{H_{\rm inf}}{M_{\rm Pl}}\right)^{2/(m-4)}H_{\rm inf},
\end{equation}
where $g_{\ast}$ is the effective number of relativistic degrees of
freedom and we have assumed the universe would evolve in the same way as
in the Einstein gravity after inflation.  If long-lived massive
particles are copiously produced at the gravitational particle
production, the reheating temperature may be significantly
 higher then the above value.  Furthermore, the decay of quasi-flat
 direction may produce a large amount of entropy to reheat the universe
 efficiently and  create 
 matter particles~\cite{Enqvist:2003qc}.

\if
At this point one may wonder if
standard particle production due to the decay of the inflaton
might work in our scenario. In such a reheating process, the inflaton decay (and
the associated energy loss of the inflaton) must be very efficient in order to
avoid the dominance of the inflaton energy density in comparison to that
of produced radiation. For this reason the inflaton must oscillate after
inflation. However, in our construction of the action in the ADM
language, the field $\phi$ is used to specify constant time hypersurfaces
and so $\phi$ and the cosmic time $t$ have a one-to-one
correspondence. Therefore, our approach cannot describe the oscillatory phase of the
inflaton and we need to resort to
gravitational reheating instead of the standard reheating scenario due to
the inflaton decay. We expect that a different action based on the $\phi$
language instead of the ADM language can accommodate the oscillatory phase
as well as the Genesis and inflationary phases.
Minimum modification would be such that one moves to the $\phi$ language
at the end of inflation to construct a $\phi$-Lagrangian that admits
the standard reheating scenario after inflation.
However, this is beyond the scope of the present paper and is left for future work.
\fi

\section{Primordial Fluctuations and Stability}

Having obtained the background evolution of our scenario,
let us investigate the nature of primordial perturbations and
stability, using the result of the generic analysis in Sec.~\ref{sec:CP}.

\subsection{Genesis Phase}

During the Genesis phase, we have
\begin{eqnarray}
&&
{\cal G}_{T}\simeq\mpl^{2},
\quad
\Sigma\simeq\frac{M_{2}^{4}}{2}f^{-2(\alpha +1) }\left(N_0^{2}a_{2}'\right)',
\nonumber\\
&&
\Theta\simeq\frac{M_{3}^{3}}{2}f^{-(2\alpha +1)}N_0a_{3}'+\eta_{4}\mpl^{2}H,
\nonumber\\
&&
{\cal G}_{A}\simeq\eta_{4}\mpl^{2},
\quad
{\cal C}\simeq\frac{\mpl^{2}}{2}(\lambda_{1}-1).
\end{eqnarray}
Obviously, the kinetic term of the tensor perturbations has the right sign, ${\cal G}_T>0$.
For large $f$, we see $\Sigma{\cal C}\gg \Theta^{2}$ (as long as ${\cal C}\neq 0$),
and hence
\begin{eqnarray}
{\cal G}_{S}\simeq \frac{{\cal G}_{T}^{2}}{{\cal C}}+3{\cal G}_{T},
\quad
{\cal H}_{S}\simeq \frac{{\cal G}_{B}^{2}}{\Sigma}.
\end{eqnarray}
This implies that ${\cal G}_{S}\simeq$ const,
while ${\cal H}_S\sim (-t)^{2(\alpha +1)}$.
The kinetic term of the curvature perturbation
has the right sign if
\begin{eqnarray}
{\cal G}_S>0\;\;\Leftrightarrow \;\;\frac{3\lambda_{1}-1}{\lambda_{1}-1}>0.
\end{eqnarray}
Thus, it is sufficient to impose
\begin{eqnarray}
\lambda_1>1.
\end{eqnarray}
(We are considering only the case with $\lambda_1>1/3$.)
Another stability condition, ${\cal H}_S>0$, is equivalent to requiring that
\begin{eqnarray}
\left(N_0^{2}a_{2}'\right)'>0.
\end{eqnarray}

Since ${\cal F}_{T}$ depends on $B_{4}$ and $B_{5}$
and these two functions are irrelevant to the background dynamics,
the condition ${\cal F}_{T}>0$ can easily be satisfied without spoiling
the Genesis background.
Suppose for simplicity that
\begin{eqnarray}
B_{4}=\frac{\beta\mpl^{2}}{2},\quad B_{5}=0,
\end{eqnarray}
where $\beta\;(>0)$ is a constant.
Then, ${\cal F}_{T}={\cal G}_{B}=\beta\mpl^{2}>0$. For the scalar perturbations we have
\begin{eqnarray}
{\cal F}_{S}&\simeq& 2\beta\mpl^{2}\left[
\frac{
M_{2}^{4}a_{2}+(2\alpha+1)M_{3}^{3}(\dot f_{0}/N_0)(N_0a_{3})'}
{(2\alpha +1)(\lambda_{1}-1)M_{2}^{4}(N_0^{2}a_{2}')'}
-\frac{1}{2}
\right]
\nonumber\\
&=&{\rm const.}
\end{eqnarray}
This can also be made positive by an appropriate choice of $a_3(N)$.
It should be noted that
if $a_3=0$ then we inevitably have ${\cal F}_S<0$; the $L_3$ term
is crucial for the stable violation of the NEC.
Note also that, if we take sufficiently small $\beta$, the sound speed
$c_s$ can be smaller than unity, which applies also to the other two phases
discussed below.

Let us move to discuss the nature of the primordial fluctuations
in the Genesis phase.
Since ${\cal G}_T\sim {\cal F}_T\sim$ const,
the tensor perturbations behave in the same way as  in the Minkowski spacetime.
Therefore, no large tensor modes are generated during the first stage of our scenario.

%%%%%%%%%%%%%%%%%%%%%%%%%
\begin{figure}[tb]
  \begin{center}
    \includegraphics[keepaspectratio=true,height=110mm]{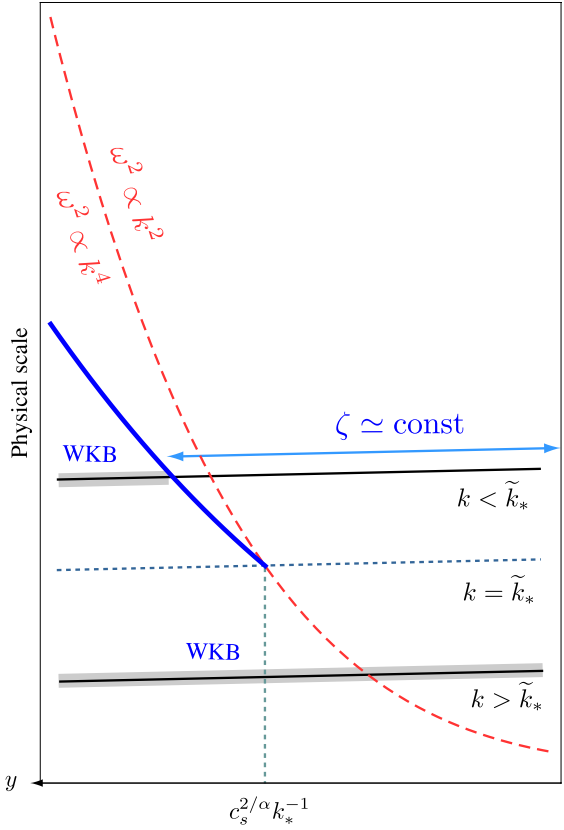}
  \end{center}
  \caption{Schematic diagram of the behavior of curvature perturbation
 in $(y, a/k)$ plane with $y$ decreasing toward the right.  In the
 region below (above) the red broken curve, $\omega^2$ is dominated by
 the term proportional to $k^4$ $(k^2)$.
Modes with $k<\widetilde{k}_*$ experience the break down of the WKB 
approximation around the point crossing the blue solid curve beyond
 which $\zeta$ is frozen,
  while modes with $k>\widetilde{k}_*$ do not.
  }%
  \label{fig:horizon.eps}
\end{figure}
%%%%%%%%%%%%%%%%%%%%%%%%%

The behavior of the curvature perturbation turns out to be more nontrivial,
as sketched in Fig.~\ref{fig:horizon.eps}.
Recalling that ${\cal G}_S\sim$ const, ${\cal F}_S\sim$ const,
and ${\cal H}_{S}\sim (-t)^{2(\alpha +1)}$,
the equation of motion for $\zeta$ in the Fourier space is of the form
\begin{eqnarray}
\frac{\D^{2}\zeta_{k}}{\D y^{2}}+\omega^{2}\zeta_{k}=0,\label{zetaeqgen}
\end{eqnarray}
where $y:= -N_0 t>0$ and
\begin{eqnarray}
\omega^{2}=c_{s}^{2}k^{2}+k_{\ast}^{2\alpha}k^{4}y^{2\alpha +2}, \label{omega}
\end{eqnarray}
with $c_{s}$ and $k_{\ast}$ being some constants.
For sufficiently large $y$, we have $\omega^{2}\approx
k_{\ast}^{2\alpha}k^{4}y^{2\alpha+2}$.  One may define the time at which
this approximation breaks down as $y_{\rm break}:=
c_s^{1/(\alpha+1)}k_{\ast}^{-\alpha/(\alpha+1)}k^{-1/(\alpha+1)}$, and
for $y\ll y_{\rm break}$ we have $\omega^2\simeq c_s^2 k^2$.

With some manipulation, it is found that
\begin{eqnarray}
\left(\frac{\D\omega/\D y}{\omega^{2}}\right)^2,\;
\left|\frac{\D^2\omega/\D y^2}{\omega^3}\right|
\lesssim
\left(\frac{\widetilde{k}_{\ast}}{k}\right)^{2\alpha/(\alpha+1)},
\end{eqnarray}
where $\widetilde{k}_{\ast}:=c_s^{-(\alpha+2)/\alpha}k_\ast$.
This implies that for the modes with $k>\widetilde{k}_\ast$ the WKB approximation
is always good in the Genesis phase,
\begin{eqnarray}
\zeta_k\propto \frac{1}{\sqrt{\omega}}\exp\left(i\int^y\omega \,\D y'\right),
\end{eqnarray}
giving $\zeta_k\propto e^{ic_sky}/\sqrt{c_sk}$ for $y\ll y_{\rm break}$.
Thus, the amplitude of those modes at late times in the Genesis phase is given by
\begin{eqnarray}
k^{3}|\zeta_{k}|^{2}\sim  \frac{k^{2}}{{\cal G}_Sc_s}\quad (k>\widetilde{k}_{\ast}).
\end{eqnarray}

For the modes with $k<\widetilde{k}_\ast$, the WKB approximation breaks
down at some time
and then the curvature perturbation freezes.
This ``horizon crossing'' occurs at $y\sim y_{\rm freeze}:= k_{\ast}^{-\alpha/(\alpha+2)}k^{-2/(\alpha+2)}$.
It can be seen that $y_{\rm freeze}>y_{\rm break}$ for $k<\widetilde{k}_\ast$,\footnote{Note
in passing that $y_{\rm break}=y_{\rm freeze}=c_s^{2/\alpha}k_\ast^{-1}$ for the $k=\widetilde{k}_\ast$ mode.
The Genesis phase could end sufficiently early so that
$-N_0t_0>c_s^{2/\alpha}k_\ast^{-1}$. If this is the case, we only need to care about
the modes with $k<\widetilde{k}_\ast$.}
which allows us to
study the freezing process by using the
solution to Eq.~(\ref{zetaeqgen}) with $\omega^2\approx k_{\ast}^{2\alpha}k^{4}y^{2\alpha+2}$.
The exact solution in this case
that matches the positive frequency WKB solution for $y\gg y_{\rm freeze}$
is given by
\begin{eqnarray}
\zeta_{k}\propto y^{1/2}H^{(1)}_{\nu}(-2\nu k_{\ast}^{\alpha}k^{2}y^{\alpha+2}),
\quad \nu:=-\frac{1}{2(\alpha +2)},
\label{smallksol}
\end{eqnarray}
where $H_\nu^{(1)}$ is the Hankel function of the first kind.
The frozen amplitude
can thus be evaluated by
taking the limit $y \ll y_{\rm freeze}$ in the solution~(\ref{smallksol}),
leading to
\begin{eqnarray}
k^{3}|\zeta_{k}|^{2} \sim \frac{k_{\ast}^{2}}{{\cal
 G}_S}\left(\frac{k}{k_{\ast}}\right)^{(3\alpha+4)/(\alpha +2)}. 
\label{frozen}
\end{eqnarray}
For $y< y_{\rm break}$, $\omega$ is dominated by the $c_sk$ term where the
solution~(\ref{smallksol}) is no longer exact.  The frozen amplitude~(\ref{frozen}),
however, is still valid even in this regime since the
solution to Eq.~(\ref{zetaeqgen}) with the effective frequency~(\ref{omega}) does not
oscillate any more and remains  constant.  Hence, the expression of the
power spectrum~(\ref{frozen}) is correct for the entire range of 
$k < \tilde{k}_\ast$.

To summarize, the power spectrum of the curvature perturbation generated during the Genesis phase
is blue and hence is suppressed on large scales.

\subsection{Inflationary Phase}

In the (de Sitter) inflationary phase, ${\cal G}_T$, ${\cal F}_T$,
${\cal G}_S$, ${\cal F}_S$, and ${\cal H}_S$ are time-independent.
We require that all those coefficients are positive during inflation
in order to avoid instabilities.

Since the quadratic action for the tensor perturbations is
essentially the same as that of generalized G-inflation,
the power spectrum of the primordial tensor perturbations is given by~\cite{Kobayashi:2011nu}
\begin{eqnarray}
{\cal P}_T=8\frac{{\cal G}_T^{1/2}}{{\cal F}_T^{3/2}}\frac{H^2_{\rm inf}}{4\pi^2}.
\end{eqnarray}

The equation of motion for the canonically normalized variable
$u_k:=\sqrt{2{\cal G}_S}a \zeta_k$ during inflation is of the form
\begin{eqnarray}
\frac{\D^2u_k}{\D \tau^2}+\left(\omega^2-\frac{2}{\tau^2}\right)u_k=0,
\end{eqnarray}
where
\begin{eqnarray}
\omega^2=c_s^2k^2+\epsilon^2 k^4\tau^2,
\end{eqnarray}
with $c_s^2={\cal F}_S/{\cal G}_S$ and
$\epsilon:=H_{\rm inf}{\cal H}_S^{1/2}/{\cal G}_S^{1/2}$ being dimensionless constants.
Here, we have introduced the conformal time $\tau\,(<0)$ defined by $a\D\tau = N\D t$.
The dispersion relations of this form have been studied
in the context of inflation, e.g., in Refs.~\cite{Martin:2002kt,Ashoorioon:2011eg}.
The positive frequency modes are given by~\cite{Ashoorioon:2011eg}
\begin{eqnarray}
u_k= \frac{e^{-\pi c_s^2/8\epsilon }W_{ic_s^2/4\epsilon,3/4}(-i \epsilon k^2\tau^2)}{(-2\epsilon k^2\tau)^{1/2}},
\end{eqnarray}
where $W_{\kappa, m}$ is the Whittaker function. Taking the limit $\tau\to 0$,
the power spectrum of the curvature perturbation can be calculated as
\begin{eqnarray}
{\cal P}_\zeta =\frac{H_{\rm inf}^2}{2{\cal G}_Sc_s^3}\frac{1}{F(c_s^2/\epsilon)},
\end{eqnarray}
where
\begin{eqnarray}
F(x):=\frac{4}{\pi}x^{-3/2}e^{\pi x/4}\left|\Gamma(5/4-ix/4)\right|^2.
\end{eqnarray}
Even in the presence of the $k^4$ term in the dispersion relation, the
power spectrum is scale-invariant in the case of exact de Sitter
inflation.
Since we have $F\to 1$ as $x\to \infty$, we recover
the result of generalized G-inflation~\cite{Kobayashi:2011nu} in the
limit $\epsilon \to 0$.  For $x\ll 1$ we have
$F\simeq(4/\pi)|\Gamma(5/4)|^2x^{-3/2}$, so that one can take the limit
$c_s^2\to 0$ smoothly to get
\begin{eqnarray}
{\cal P}_\zeta\to \frac{\pi H_{\rm inf}^2}{8{\cal G}_S|\Gamma(5/4)|^2\epsilon^{3/2}}.
\end{eqnarray}

We have approximated the inflationary phase as exact de Sitter.
If we consider a background slightly different from de Sitter
by incorporating weak time dependence in $f$,
we would be able to obtain a tilted spectrum of $\zeta$.

\subsection{Graceful Exit}

After inflation, we have ${\cal G}_T\simeq \mpl^2$, ${\cal F}_T=\beta\mpl^2$,
\begin{eqnarray}
{\cal F}_{S}&\simeq&\beta \mpl^2
\frac{-\lambda_1+1+\ell m/2}{\lambda_1-1+\ell},
\\
{\cal G}_{S}&\simeq&
\mpl^2\frac{3\lambda_1-1}{\lambda_1-1+\ell},
\\
{\cal H}_{S}&\simeq&
\frac{\beta^2\mpl^2}{H^2}\frac{\lambda_1-1}{3\lambda_1-1}\frac{\ell}{\lambda_1-1+\ell},
\end{eqnarray}
where to simplify the expression we introduced
\begin{eqnarray}
\ell :=-\frac{4}{3}\frac{(N_{\rm e}a_2)'}{(N_{\rm e}^2a_2')'}.
\end{eqnarray}
Recalling that we have been imposing $\lambda_1>1$,
all of these coefficients are positive provided that
$\ell m>2(\lambda_1-1)$. This condition can be written equivalently as
\begin{eqnarray}
\frac{N_{\rm e}a_2'}{(N_{\rm e}^2a_2')'}<-\frac{1}{2}\left(\lambda_1-1\right)\;(<0).\label{GE:st}
\end{eqnarray}

\section{A Concrete Example}

%%%%%%%%%%%%%%%%%%%%%%%%%
\begin{figure}[tb]
  \begin{center}
    \includegraphics[keepaspectratio=true,height=100mm]{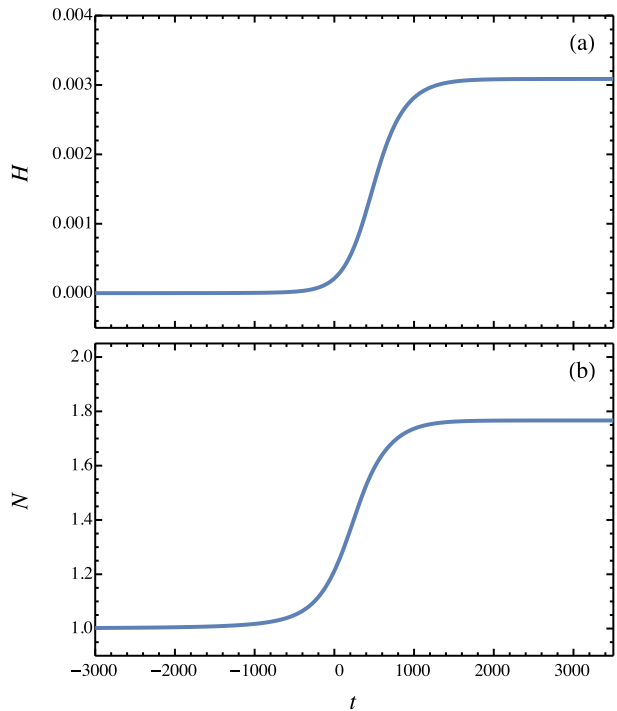}
  \end{center}
  \caption{The background evolution of (a)
  the Hubble parameter $H$  and (b) the lapse function $N$ 
  around the Genesis-de Sitter transition.
  }%
  \label{fig:HN.eps}
\end{figure}
%%%%%%%%%%%%%%%%%%%%%%%%%
\begin{figure}[tb]
  \begin{center}
    \includegraphics[keepaspectratio=true,height=100mm]{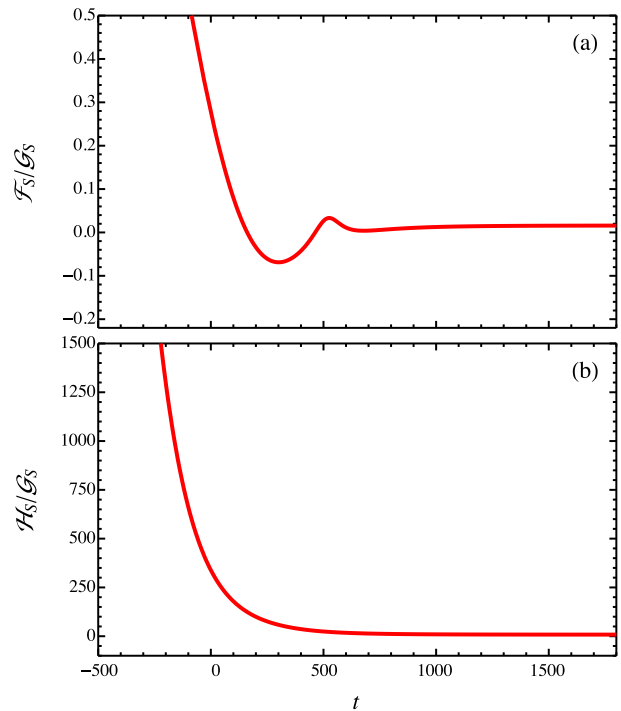}
  \end{center}
  \caption{(a) The sound speed squared,
  ${\cal F}_{S}/{\cal G}_{S}$,  and (b) 
the coefficient of $k^{4}$ (divided by ${\cal G}_{S}$) 
  around the Genesis-de Sitter transition.
  }%
  \label{fig:stability.eps}
\end{figure}
%%%%%%%%%%%%%%%%%%%%%%%%%

Let us provide a concrete Lagrangian
exhibiting the Genesis-de Sitter transition.
The Lagrangian is characterized by
\begin{eqnarray}
a_2=-\frac{1}{N^2}+\frac{N_{0}^2}{3N^4},
\quad
a_3=\frac{\gamma}{N^3},\label{example}
\end{eqnarray}
where $N_{0}\,(>0)$ and $\gamma\, (>0)$ are constants.
We take $a_4=a_5=0$, $B_{4}=\mpl^{2}/2$, and $B_{5}=0$.
We also take $\lambda_{1}>1$ to guarantee the stability.
This corresponds to the ($\lambda_1>1$ generalization of the) unitary gauge description of the
Lagrangian considered in Ref.~\cite{Creminelli:2010ba}.
In the Genesis stage we have
\begin{eqnarray}
&&
N=N_{0},
\\
&&
\hat p =-\left[\frac{2M_{2}^{4}}{3N_{0}^{2}}
+(2\alpha+1)\frac{\gamma}{N_{0}^{4}}M_{3}^{3}|\dot f_{0}|\right]<0.
\end{eqnarray}
Since $\lambda_{1}>1$ and $(N_0a_{2}')'=2/N_{0}^{2}>0$,
we see that ${\cal G}_{S}>0$ and ${\cal H}_{S}>0$.
We also see that
\begin{eqnarray}
\frac{{\cal F}_{S}}{\mpl^{2}}=\frac{2}{\lambda_{1}-1}
\left[\frac{\gamma M_{3}^{3}|\dot f_{0}|}{M_{2}^{4}N_{0}^{2}}
-\frac{1}{3(2\alpha +1)}\right] -1,
\end{eqnarray}
and hence
it is easy to satisfy ${\cal F}_{S}>0$ during the Genesis phase
by choosing the parameters appropriately.

A numerical example of the Genesis-de Sitter transition
is illustrated in Figs.~\ref{fig:HN.eps} and~\ref{fig:stability.eps}.
Our numerical calculation was performed as follows: we solve the evolution equations
${\cal P}=0$ and $\D{\cal E}/\D t=0$ with
initial data $(H, N)$ satisfying ${\cal E}=0$, and confirm that
the constraint ${\cal E}=0$ is satisfied at each time step.
In the numerical calculation, the parameters are given by
$\mpl=M_{2}=M_{3}=1$,
$\alpha =1$, $\lambda_{1}=1+10^{-3}$, $N_{0}=1$, and $\gamma =10$.
The function $f(t)$ is taken to be
\begin{eqnarray}
f=\frac{\dot f_{0}}{2}\left[t-\frac{\ln(2\cosh(st))}{s}\right]+f_{1},\label{f:ex1}
\end{eqnarray}
with $\dot f_{0}=-10^{-1}$, $f_{1}=10$, and $s=2\times 10^{-3}$.
The background evolution is shown in Fig.~\ref{fig:HN.eps}.
The evolution of the sound speed squared, ${\cal F}_{S}/{\cal G}_{S}$,
and the coefficient of $k^{4}$ in the dispersion relation is shown in Fig.~\ref{fig:stability.eps}.
As pointed out in Ref.~\cite{Pirtskhalava:2014esa},
$c_{s}^{2}$ flips the sign at the transition.
The sound speed squared is positive except in this finite period.
During the Genesis and subsequent de Sitter phases
we have ${\cal G}_S>0$ and ${\cal H}_S>0$, and therefore we may conclude that
this model is stable.

%%%%%%%%%%%%%%%%%%%%%%%%%
\begin{figure}[tb]
  \begin{center}
    \includegraphics[keepaspectratio=true,height=100mm]{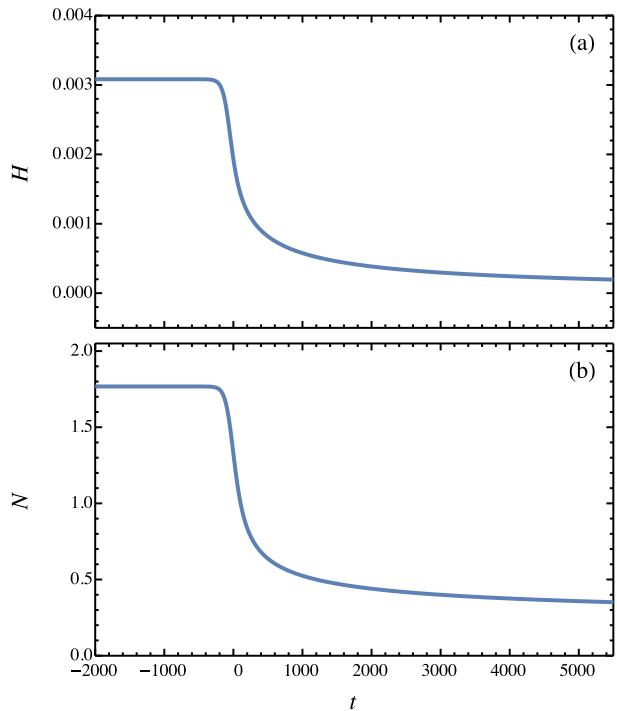}
  \end{center}
  \caption{The background evolution of (a)
  the Hubble parameter $H$  and (b) the lapse function $N$ 
  around the end of inflation.
  }%
  \label{fig:HN-dS-R.eps}
\end{figure}
%%%%%%%%%%%%%%%%%%%%%%%%%
\begin{figure}[tb]
  \begin{center}
    \includegraphics[keepaspectratio=true,height=100mm]{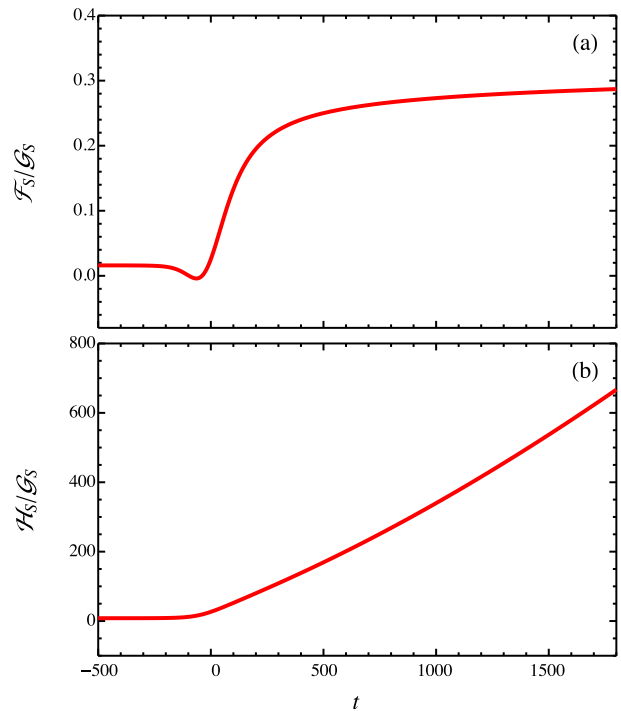}
  \end{center}
  \caption{(a) The sound speed squared,
  ${\cal F}_{S}/{\cal G}_{S}$,  and (b)
the coefficient of $k^{4}$ (divided by ${\cal G}_{S}$) 
  around the end of inflation.
  }%
  \label{fig:stability-dS-R.eps}
\end{figure}
%%%%%%%%%%%%%%%%%%%%%%%%%

Although we have thus obtained the stable example of the Genesis-de Sitter transition,
the simple example~(\ref{example}) is not completely satisfactory
if one would want successful gravitational reheating.
Indeed, the condition~(\ref{exit:c1}) implies that $x:=(N_{\rm e}/N_0)^2<1$, but
$m-4=-2x/(1-x)<0$ for such $x$.
This problem can be evaded easily by the following small deformation of $a_2$:
\begin{eqnarray}
a_2=-\frac{1}{N^2}+\frac{1+5\Delta^2}{3}\frac{N_0^2}{N^4}-\Delta^2\frac{N_0^4}{N^6},
\end{eqnarray}
where $\Delta$ is a parameter smaller than $1/5$.
The condition~(\ref{exit:c1}) now reads
$(1-x)(x-5\Delta^2)>0$, i.e., $5\Delta^2<x<1$, while
\begin{eqnarray}
m-4=\frac{2(\Delta +x)(\Delta-x)}{(1-x)(x-5\Delta^2)}
\end{eqnarray}
is positive for $5\Delta^2<x<\Delta$.
The stability condition further restricts
the allowed ranges of $x$ and $\Delta$.
The necessary condition for stability is $N_{\rm e}a_2'/(N_{\rm e}^2a_2')'<0$ [see Eq.~(\ref{GE:st})].
This translates to
$1+5\Delta^2-\sqrt{1-5\Delta^2+25\Delta^4}<x<\Delta<(4-\sqrt{11})/5\simeq 0.137$,
leading to $m<24/5$ = 4.8.
Note that the small deformation of $a_2$ with $\Delta \lesssim 0.1$
does not change the background and perturbation dynamics of the Genesis and inflationary phases.

To illustrate the final stage of inflation, let us take
\begin{eqnarray}
f=\left\{f_1^{\alpha +1}
+\frac{v}{2}\left[t+\frac{\ln\left(2\cosh(s' t)\right)}{s'}\right]\right\}^{1/(\alpha +1)},\label{f:ex2}
\end{eqnarray}
where the origin of time is shifted so that the end of inflation is given by $t\sim 0$.
In the numerical plots presented in Figs.~\ref{fig:HN-dS-R.eps} and~\ref{fig:stability-dS-R.eps},
the parameters are given by
$s'=10^{-2}$, $v=6$, and $\Delta = 0.05$, while the other parameters are taken to be the same as
the previous example of the Genesis-de Sitter transition.
It is found that $m\sim 4.5>4$.
Again, we see that $c_s^2<0$ in the finite period around the transition.
However, ${\cal G}_S$ and ${\cal H}_S$ remain positive all through the inflation and subsequent stages.

\section{Discussion and conclusion}

In this paper, we have
introduced a generic description of Galilean Genesis
in terms of the ADM Lagrangian and
constructed a concrete realization of
inflation preceded by Galilean Genesis, {\it i.e.},
the scenario
in which the universe starts from Minkowski spacetime in the asymptotic
past and is connected smoothly to the inflationary phase followed by the graceful
exit. Our model utilizes the recent extension of the Horndeski theory, which
has the same number of propagating degrees of
freedom as the Horndeski theory and thus can avoid Ostrogradski instabilities.
This approach allows us to cover
the background and perturbation evolution in all the three
phases with the same single Lagrangian,
as opposed to the effective field theory approach.
In our scenario, the sound
speed squared during the transition from the Genesis phase to inflation
becomes negative for a short period. However, thanks to the nonlinear
dispersion relation arising from the fourth-order derivative term in the
quadratic action, modes with higher momenta are stable and
the growth rate of perturbations with smaller momenta is finite and
under control. It should also be noted that the sound speed of the
primordial perturbations can be smaller than unity by choosing the parameter of
the model appropriately.

%In the inflationary scenario completed with Galilean Genesis to the past,
%observable primordial tensor perturbations can be produced, contrary to the
%original Genesis scenario as an alternative to
%inflation~\cite{Pirtskhalava:2014esa}.

Although we have constructed our inflation model in order to resolve the
initial singularity and possible trans-Planckian problems by
incorporating Galilean Genesis phase before inflation, we could make use
of our model to realize the original Galilean Genesis scenario, which is  an
alternative to inflationary cosmology, simply by taking vanishingly
short period of inflation there.
As discussed
in the Appendix, the sound speed squared becomes negative at the transition
also in this case, but the instabilities are relevant
only for small $k$ modes thanks to the $k^4$ term in the dispersion relation.
Thus,
the transition from the Genesis phase to the reheating stage is described
in a healthy and controllable manner. 

In fact, it would be fair to say that such a cosmology works quite well
among the proposed alternatives to inflation, because, in contrast with
the bouncing cosmology, in which all the would-be decaying modes in the
expanding universe such as vector fluctuations and spatial anisotropy
severely increase in an undesirable manner, the Genesis solution is an
attractor and generation of nearly scale-invariant curvature
perturbation is also possible with an appropriate choice of model
parameters \cite{Nishi:2015pta}.  Since no first-order tensor
perturbation is generated in this type of scenarios, detection of tensor
perturbation with its amplitude larger than $10^{-10}$ would be a
smoking gun of inflation.

%--- Acknowledgements ---%--- Acknowledgements ---%--- Acknowledgements ---%
\acknowledgments 
This work was supported in part by the JSPS Grant-in-Aid for Scientific
Research Nos.~24740161 (T.K.), 25287054 and 26610062 (M.Y.), 23340058
and 15H02082 (J.Y.).
%--- Acknowledgements ---%--- Acknowledgements ---%--- Acknowledgements ---%

%-------------------------------------------------------------------%

\appendix
\section{Matching Genesis to the Reheating Phase}

%%%%%%%%%%%%%%%%%%%%%%%%%
\begin{figure}[tb]
  \begin{center}
    \includegraphics[keepaspectratio=true,height=100mm]{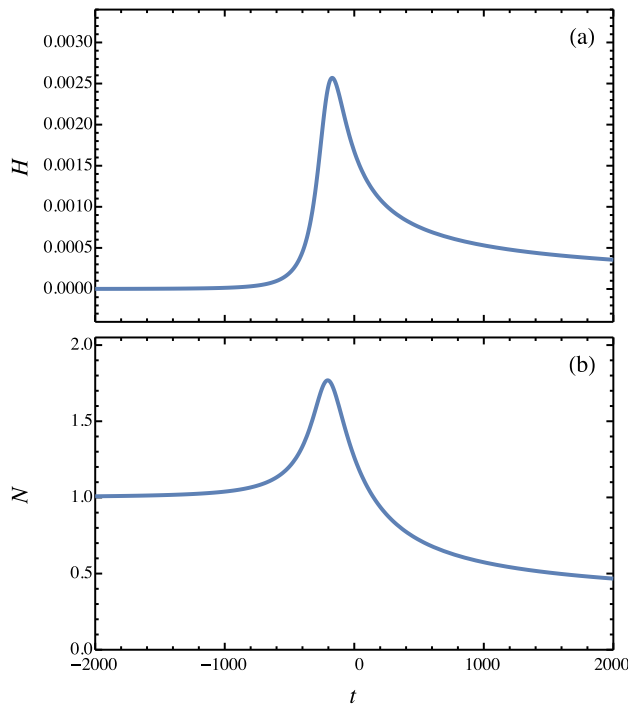}
  \end{center}
  \caption{The background evolution of (a)
  the Hubble parameter $H$  and (b) the lapse function $N$ 
  around the Genesis-reheating transition.
  }%
  \label{fig:HN-G-R.eps}
\end{figure}
%%%%%%%%%%%%%%%%%%%%%%%%%
\begin{figure}[tb]
  \begin{center}
    \includegraphics[keepaspectratio=true,height=100mm]{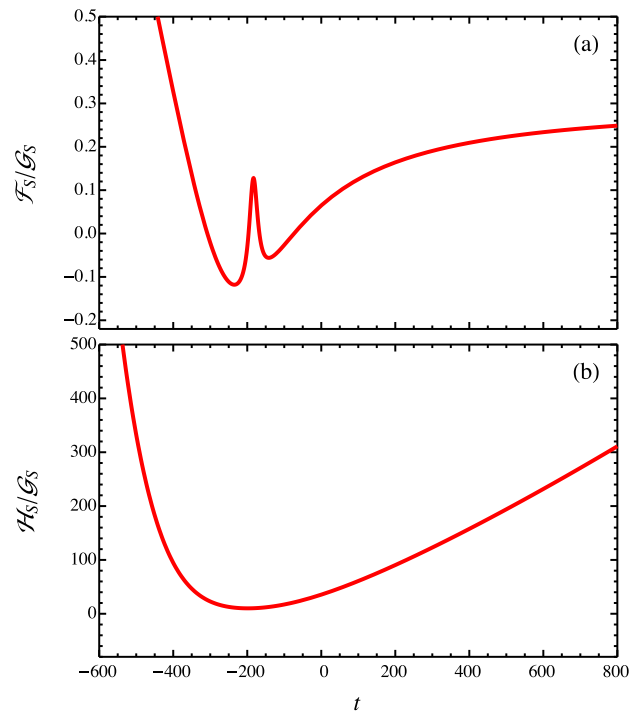}
  \end{center}
  \caption{The sound speed squared
  ${\cal F}_{S}/{\cal G}_{S}$ (a) and the coefficient of $k^{4}$ (divided by ${\cal G}_{S}$) (b)
  around the Genesis-reheating transition.
  }%
  \label{fig:stability-G-R.eps}
\end{figure}
%%%%%%%%%%%%%%%%%%%%%%%%%

In the main text, we consider the scenario in which Galilean Genesis
is followed by inflation. In this appendix, we will go back to the original motivation
of Galilean Genesis and study how we can match smoothly the Genesis phase
to the reheating phase.
Our approach based on the ADM Lagrangian is quite useful in
analyzing such a situation as well.

It is now obvious that by taking
\begin{eqnarray}
f\sim
\begin{cases} 
|t|,  & \mbox{for}\;\; t<0 \\
t^{1/(\alpha +1)}, & \mbox{for }\;\; t>0 
\end{cases}\;,
\end{eqnarray}
and gluing the two functions smoothly at around $t=0$,
one can describe the Genesis-reheating transition.
As a concrete example, we glue $f\approx 0.1 (-t)$ and $f\approx (6t)^{1/2}$ smoothly at around $t=0$
and perform a numerical calculation as shown in Figs.~\ref{fig:HN-G-R.eps} and~\ref{fig:stability-G-R.eps}.
The other parameters are the same as those taken in the main text.
As is expected, the numerical result here is much the same as
the case where a duration of the intermediate inflationary phase
is taken to be very short.
In particular, $c_s^2$ becomes negative at the Genesis-reheating transition.
The model is nevertheless stable since the conditions ${\cal G}_S>0$ and ${\cal H}_S>0$ remain satisfied.

%-------------------------------------------------------------------%

%---------   References   ---------%

\end{document}